# Response of a multi-wavelength laser to single-sideband optical injection

Shahab Abdollahi, Pablo Marin-Palomo, and Martin Virte

*Abstract*— Single-mode semiconductor lasers subject to optical injection have been shown to trigger a wide range of dynamical behavior from injection locking to chaos. Multi-wavelength lasers add even more degrees of freedom and complexity to the dynamical repertoire potentially unlocking new functionalities for applications ranging from THz generation and processing to all-optical memories. In particular, leveraging the inherent mode coupling in multi-wavelength lasers, spectral multiplication over a THz range of an injected optical signal has been shown. While most of the research on optical injection has been focused on single-mode semiconductor lasers, the dynamical behavior of multi-wavelength lasers, particularly when subjected to injection of amplitude-modulated signals remains vastly unexplored. In this work, we numerically and experimentally investigate the response of an on-chip dual-wavelength laser subject to the optical injection of a single-sideband signal around one of the modes of the laser. Our findings show an asymmetric power evolution of the sidebands appearing around both the injected and un-injected modes with respect to the modulation frequency. The power and bandwidth of the sideband signals strongly depend on the resonance frequency produced by the interference between the cavity mode and the injection, which can be tailored by twerking the strength and the detuning of the injection. The outcomes of our numerical investigations, based on rate equations, align closely with the experimental results highlighting the influence of key injection and laser parameters.

## I. INTRODUCTION

Semiconductor lasers subjected to optical injection exhibit a wide range of dynamical behaviors, including stable locking, periodic oscillations, and chaos [1–4]. The utility of injection locking in improving laser performances, such as reducing linewidth, suppressing mode hopping, and enhancing modulation bandwidth, has been extensively explored [5–7]. Outside the locking range, different types of dynamics emerge which have been used in various applications such as THz signal generation, chaotic communication, and random bit generation [8–10]. The dynamics of optically injected semiconductor lasers become considerably more intricate when exposed to multiple injection beams or broad-band amplitude-modulated signals such as optical frequency combs [11,12]. In particular, a single-mode semiconductor laser, injected with a frequency comb has been observed to exhibit a different dynamical behavior compared to the one in the case of single-frequency injection [13]. Troger et al. [14] conducted a theoretical and experimental investigation into the dynamics of a slave laser under the influence of multiple external injection beams, revealing complex behavior compared to single-beam injection. Al-Hosiny et al. [15] performed a more extensive investigation and observed a secondary locking region in a semiconductor laser optically injected by two master lasers, indicating additional dynamics beyond conventional single-beam injection scenarios. While single-mode semiconductor lasers are effective in delivering a consistent beam of light, they struggle to accommodate the diverse wavelength needs or simultaneous operation at multiple wavelengths required by many applications. This is precisely where multi-wavelength lasers (MWL) step in. A MWL is, in essence, a multimode laser whose emission is controlled and limited to a selected number of wavelengths [16]. This key capability opens the horizon of applications to various domains ranging from spectroscopy to optical communications [17–19]. When optical injection is applied to MWLs, they can serve as low-noise THz sources [20] or facilitate all-optical THz signal processing [21,22] and all-optical memory [23,24]. In addition, optical injection into a MWL triggers new complex dynamics, where each additional wavelength introduces new degrees of freedom [25–27]. This phenomenon has been explored both experimentally and numerically, as reported by Heinricht et al. [28], who investigated the locking bistability of a two-color MWL under single-mode optical injection. The injection of modulated signals into MWLs has garnered attention due to its potential to enhance the complexity of the laser's behavior. Notably, injecting chaotic signals into cascaded-coupled semiconductor lasers has emerged as a possible solution to make the chaotic behavior more complex and to improve the dynamical features for chaos-based applications [29–31]. We also recently demonstrated the spectral multiplication of a narrowband frequency comb over a large frequency offset by injecting a narrow frequency comb into an integrated MWL [32]. Despite the demonstrations of MWL dynamics under various injection schemes, there remains a notable gap in the understanding of the response of MWL to modulated optical injection in particular when it comes to the bandwidth and power characteristics of the regenerated and spectrally multiplied signals emitted, respectively, around the injected and un-injected modes of the slave laser.

In this work, we experimentally and numerically analyze the effect of modulated optical injection into a MWL. The optical injection consists of a single-sideband (SSB) signal which is emulated by combining two highly coherent light beams separated by a certain frequency offset, referred to henceforth as modulation frequency. The injection of such a single-sideband signal allows us to investigate asymmetries in the output of the slave laser. Our findings reveal an asymmetric, with respect to the modulation frequency, of the power of the sidebands appearing around the injected and the un-injected modes of the MWL. Notably, the signal emitted around the injected mode, i.e., the regenerated signal, exhibits characteristics of single-sideband modulation, while the signal emerging around the un-injected mode, termed as multiplied signal, experiences dual-sideband modulation. Moreover, we observe a strong correlation between the power of the sidebands of both regenerated and multiplied signals and the cavity resonance and relaxation oscillation frequency of the MWL under injection. These insights not only represent a step forward in the fundamental understanding of mode coupling in MWL, but they can also be extended to broader types of optical injection, such as data injection for wavelength conversion [33], optical frequency comb injection for comb spectral multiplication [32], or THz generation due to the large spectral separation between the modes of our MWL [16].

The following sections of this paper are organized as follows: Section II describes the experimental setup for the injection of the SSB signal into the MWL, Section III presents and discusses the experimental results, focusing on the observed dynamics and power evolution of the sidebands, Section IV introduces the numerical model and simulation and compares the theoretical predictions with experimental findings, and Section V concludes the paper summarizing the key results.

## II. EXPERIMENTAL SETUP FOR THE INJECTION OF SSB SIGNAL

The fiber-based setup used for our experiments is illustrated in Fig. 1(a). The slave laser is a dual-cavity InP distributed Bragg reflector (DBR) laser with a single semiconductor optical amplifier (SOA) serving as the common gain medium. The dual cavity laser is established through the sequential arrangement of two DBR structures on one side, and a two-port broadband reflector (BR) on the other side [32,34]. The pitch of the DBRs is set to obtain a spectral separation of 10 nm between the two reflectivity bands. This enables our MWL to emit at wavelengths separated by more than a THz, see Fig. 1(b). During our experiments, the temperature of the chip is set to 22 ºC. The MWL is biased at $J = 50$ mA which is 2.2 times its threshold current ($J_{th} = 21$ mA). In this configuration, the free-running laser emits at $\lambda_{2,fr} = 1547.614$ nm while the other longitudinal modes are strongly suppressed by more than 40 dB, see Fig. 1(b).

The SSB signal is emulated by merging two independent low-noise tunable external-cavity lasers (Keysight N7776C), labeled $M_1$ and $M_2$, featuring a short-term linewidth of approximately 10 kHz. Laser $M_1$ serves as the carrier (C), while $M_2$ corresponds to the sideband (SB), and the modulation frequency is defined as $f_m = f_{SB} - f_C$, see Fig. 1(c) To precisely track the modulation frequency, we impinge the merged lasers into a photodiode (PD$_2$). The power difference $P_m = P_{SB}/P_C$ between the two master lasers is controlled with a variable optical attenuator (VOA$_1$) and set to $-10$ dB during the experiment. While the SSB signal is generated by merging two independent lasers, limiting the coherence of the SSB signal, we do not expect to have a qualitative impact on the response of the MWL. This choice enables us to explore positive and negative modulation frequencies separately and their effects on the response of our MWL. The SSB signal is injected into the cavity of the MWL via a lensed fiber (LF) and a fiber optical circulator (FOC) to avoid back reflections. We use a polarization controller (PC) to align the polarization of the injection to that of the slave laser, and VOA$_2$ to adjust the injection strength $\kappa_{inj} = P_{in}/P_{out}$, where $P_{in}$ and $P_{out}$ are the power to and power from the MWL measured after the lensed fiber. We inject the SSB signal detuned in frequency by $\Delta f_{inj} = f_C - f_{1,fr}$ from the suppressed mode ($\lambda_{1,fr} = 1537.106$ nm), while the mode at $\lambda_{2,fr}$ remains un-injected [35,36]. Note that the frequencies $f_{1,fr}$ and $f_{2,fr}$ of the injected and un-injected modes, respectively, are measured before injection. After injection, the cavity modes shift to $f_{1,cav}$ and $f_{2,cav}$ due to the anti-guidance effect [37,38].

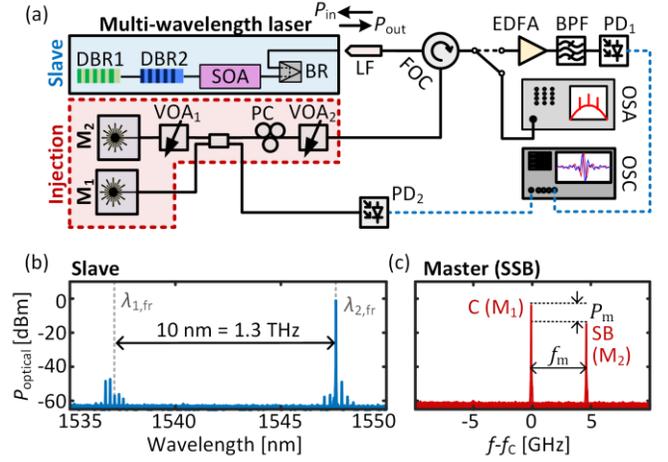

**Fig. 1.** (a) Setup for SSB signal injection into the MWL (b) Optical spectrum of the MWL without injection measured after the LF. Injected and un-injected modes are identified by vertical dashed lines at $\lambda_{1,fr}$ and $\lambda_{2,fr}$, respectively. (c) Optical spectrum of the injected SSB signal consisting of a carrier (C) and sideband (SB) measured before the LF.

The output of the MWL is either sent to an optical spectrum analyzer (OSA, 5 MHz RBW) or a high-speed oscilloscope (OSC, Lecroy, 20 Gs/s, 8 GHz bandwidth) after amplification, using an erbium-doped fiber amplifier (EDFA) to compensate for coupling losses, and detection with a photodiode (PD$_1$, 12 GHz bandwidth). A bandpass filter (BPF) is employed to select one of the modes under investigation.

## III. EXPERIMENTAL RESULTS AND DISCUSSION

We study the response of the MWL when subject to injection with an SSB signal. Figure. 2(a) shows the optical spectrum, centered around $f_C$, of the injected mode when the SSB signal is injected around $f_{1,fr}$ with the detuning and the injection strength set to $\Delta f_{inj} = -1.2$ GHz and $\kappa_{inj} = 2.1$ dB, and the modulation frequency to $f_m = 2.7$ GHz. The modes labeled $C_{reg}$ and SB$_{reg}$ correspond to the regenerated amplified SSB signal. In addition, we observe the generation of a nonlinear sideband SB$_{NL}$ at a frequency offset from $C_{reg}$ corresponding to $f_m$, appearing from the interaction between the SB$_{reg}$ and $C_{reg}$. We further observe that the signal is multiplied to the un-injected mode, 1.3 THz away from the injected mode, leading to the emergence of sidebands on the left ($\text{SB}_{\text{mult,L}}$) and right ($\text{SB}_{\text{mult,R}}$) side of the un-injected mode, see Fig. 2(b). The spectral multiplication of the injected signal around the un-injected mode can be explained by the amplitude modulation experienced by the carrier population of the laser [32]. In Fig. 2(c) and (d), we show the power of the sidebands emerging around both the injected and un-injected modes for a fixed detuning of $\Delta f_{\text{inj}} = -1.2$ GHz and injection strength of $\kappa_{\text{inj}} = 2.1$ dB, while sweeping $f_{\text{m}}$ from $-5$ GHz to 5 GHz. Here, we observe that the amplitude of the multiplied sidebands strongly depends on $f_{\text{m}}$. For $f_{\text{m}}$ values near 1.6 GHz, the sideband (SB) is injected close to the cavity mode at $f_{1,\text{cav}}$, leading to a gain boost of both the $\text{SB}_{\text{reg}}$ and $\text{SB}_{\text{NL}}$. Such gain increase, labelled as I in Fig. 2(c) and subsequent figures, occurs when $f_{\text{m}} \approx f_{\text{res}}$, with the resonance frequency being $f_{\text{res}} = f_{1,\text{cav}} - f_{\text{C}}$.

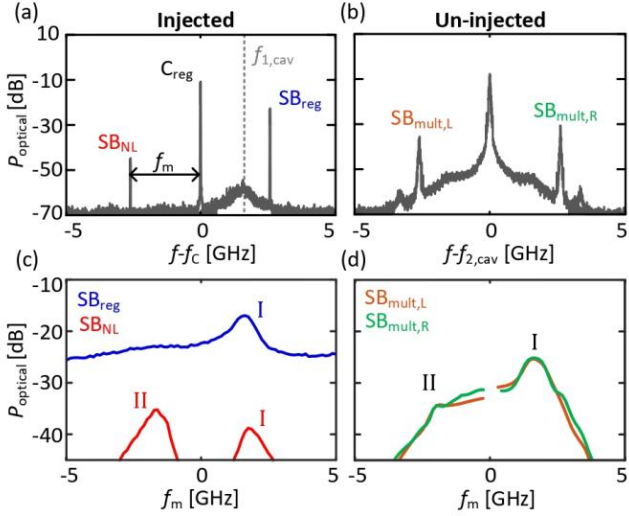

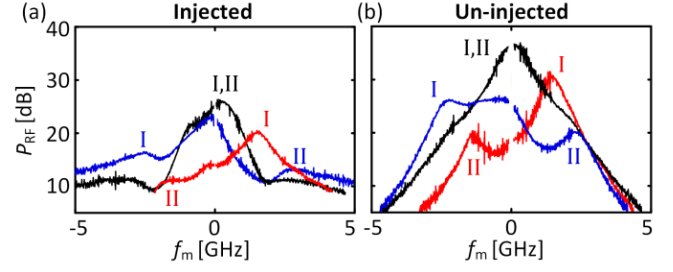

**Fig. 2.** Optical spectrum of the regenerated (a) and multiplied (b) signals around the injected and un-injected modes, respectively for $\Delta f_{\text{inj}} = -1.2$ GHz, $\kappa_{\text{inj}} = 2.1$ dB, and $f_{\text{m}} = 2.7$ GHz. (c) Optical power of regenerated sidebands $\text{SB}_{\text{reg}}$ (dark blue trace) and $\text{SB}_{\text{NL}}$ (red trace) emerging around the injected mode and (d) of the multiplied sidebands $\text{SB}_{\text{mult,R}}$ (green trace) and $\text{SB}_{\text{mult,L}}$ (orange trace) emerging around the un-injected mode.

On the other hand, when $f_{\text{m}} \approx -f_{\text{res}}$, labeled as II in Fig. 2(c) and subsequent figures, the $\text{SB}_{\text{NL}}$ emerges close to the cavity mode and experiences an increase in power that is larger than that for $f_{\text{m}} \approx f_{\text{res}}$. For negative $f_{\text{m}}$ values, $\text{SB}_{\text{reg}}$ maintains a rather constant power level until approximately $-2$ GHz, which is related to the relaxation oscillation (RO) frequency of the injected MWL. Thus, around the injected mode, we observe a clear asymmetry in the evolution of the sideband power as a function of modulation frequency. In addition, besides the power difference between $\text{SB}_{\text{reg}}$ and $\text{SB}_{\text{NL}}$, their dependence with respect to $f_{\text{m}}$ is fundamentally different as the $\text{SB}_{\text{NL}}$ emerges from the interaction between the signal's carrier and sideband.

**Fig. 3.** RF power of the beatnote at frequency $f_{\text{m}}$, generated from the beating of (a) $\text{C}_{\text{reg}}$, $\text{SB}_{\text{reg}}$ and $\text{SB}_{\text{NL}}$, and (b) the un-injected mode, at $f_{2,\text{cav}}$, with $\text{SB}_{\text{mult,L}}$ and $\text{SB}_{\text{mult,R}}$. The red, black, and blue traces correspond to $\Delta f_{\text{inj}}$ values of $-1.2$ GHz, 0 GHz, and 1.2 GHz, respectively with $\kappa_{\text{inj}}$ fixed at 2.1 dB.

In contrast, the power of the multiplied left ($\text{SB}_{\text{mult,L}}$) and right ($\text{SB}_{\text{mult,R}}$) sidebands emerging around the un-injected mode follows the same trend with only minor fluctuation for low $f_{\text{m}}$ values, see Fig. 2(d). This is particularly visible for $f_{\text{m}} \approx f_{\text{res}} = 1.6$ GHz where the power of both sidebands is increased equally, i.e., the power of both multiplied sidebands is similarly influenced by the cavity resonance frequency, $f_{\text{res}}$, of the injected mode. However, we also observe a clear asymmetry in the sideband power as a function of $f_{\text{m}}$. Naturally, adjusting $\Delta f_{\text{inj}}$ changes $f_{\text{res}}$ thereby resulting in a different $f_{\text{m}}$ value at which the sideband powers are amplified. Qualitatively, we obtain a similar behavior in the case that both injected and un-injected modes have a similar power, instead of the injected mode being suppressed. In this case, however, we are constrained to low injection strengths, to avoid the suppression of the un-injected mode. In addition, while the multiplied sidebands feature higher power values, we do observe dynamical behavior, refer to Section S2 in Supplementary Materials.

In the following, we explore the role of injection parameters such as detuning and injection strength, in shaping the power of the sidebands of both regenerated and multiplied signals. With this investigation, we clarify the impact of the resonance frequency and the RO on the strength of the regenerated and multiplied signals. In the first experiment, the injection strength is kept constant at $\kappa_{\text{inj}} = 2.1$ dB, to ensure that enough power is injected into the MWL. Meanwhile, the detuning $\Delta f_{\text{inj}}$ is set to three different values: $-1.2$ GHz, 0 GHz, and 1.2 GHz, in order to cover different detuning scenarios. We sweep the modulation frequency of the injection from $f_{\text{m}} = -5$ GHz to $f_{\text{m}} = 5$ GHz, to scan a fair range of frequency offsets around the injected mode. Here, we use the oscilloscope, rather than the OSA, to capture, in a single measurement, the MWL output throughout the entire sweep. We slice the recording resulting in an $f_{\text{m}}$ change of 6.2 MHz per slice, perform the Fourier transform, and measure the RF power, $P_{\text{RF}}$, for each slice at the Fourier frequency corresponding to $f_{\text{m}}$, see Fig. 3.

In the case of the injected signal, due to the low power of $\text{SB}_{\text{NL}}$, $P_{\text{RF}}$ solely corresponds to the power of the beating between $\text{C}_{\text{reg}}$ and $\text{SB}_{\text{reg}}$. While for the un-injected signal, $P_{\text{RF}}$ is obtained from the beatings between un-injected mode at $f_{2,\text{cav}}$ with both $SB_{mul,L}$ and $SB_{mul,R}$. Since the power of both $C_{reg}$ and un-injected mode remains constant during the measurement, the RF power, shown in Fig. 3, provides an appropriate overview of the power of the sidebands.

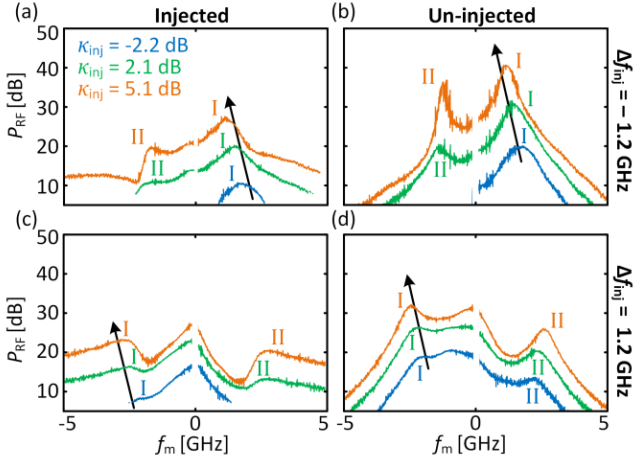

**Fig. 4.** RF power of the beatnote at frequency $f_m$ generated around the injected (a) and un-injected (b) modes for $\Delta f_{inj} = -1.2$ GHz and $\kappa_{inj} = -2.2$ dB (blue trace), $\kappa_{inj} = 2.1$ dB (green trace), and $\kappa_{inj} = 5.1$ dB (orange trace). RF power of the beatnote at frequency $f_m$ generated around the injected (c) and un-injected (d) modes for $\Delta f_{inj} = 1.2$ GHz and the same injection strength values. For the injected mode the RF power is generated from the interaction between $C_{reg}$, $SB_{reg}$ and $SB_{NL}$ and for the un-injected mode it is the beating between the un-injected mode, at $f_{2,cav}$, and $SB_{mult,L}$ and $SB_{mult,R}$.

For the negative detuning, red trace in Fig. 3(a), $f_{res} \approx f_{RO}$, where $f_{RO}$ is measured to be approximately 1.5 GHz at $J = 50$ mA. As a result, the sidebands emerging around the injected mode are significantly enhanced when $f_m \approx f_{res}$ (case I). It is worth noting that while the optical injection enhances the RO frequency of the laser, this increase is negligible compared to the shift of the cavity mode frequency caused by the anti-guidance effect [39]. Similarly, due to the carrier density modulation and intermodal cross-coupling, the sidebands of the multiplied signal emerging around the un-injected mode experience a gain boost as indicated with I in the red trace of Fig. 3(b). The black traces in Fig. 3(a) and (b), shows the RF power of the sidebands for a detuning $\Delta f_{inj} = 0$ GHz. In this scenario, the carrier is injected close to the cavity mode at $f_{1,cav}$ while the RO frequency is approximately the same, resulting in a gain boost at around $f_m = 0$ GHz. Finally, the blue lines in Fig. 3(a) and (b) correspond to $\Delta f_{inj} = 1.2$ GHz. In this scenario $f_{1,cav}$ is shifted towards lower values increasing $f_{res}$, resulting in a peak of the gain at around $f_m = -2.1$ GHz (I) and at $f_m = 2.1$ GHz (II), along with a discernible peak around $f_{RO}$ of the MWL.

In a second experiment, we investigate three different $\kappa_{inj}$ values for a positive $\Delta f_{inj} = 1.2$ GHz and a negative $\Delta f_{inj} = -1.2$ GHz detuning values. In Fig. 4, we show the corresponding $P_{RF}$ associated to the power of the sidebands when sweeping the modulation frequency from $f_m = -5$ GHz to $f_m = 5$ GHz. For $\Delta f_{inj} = -1.2$ GHz, see Fig. 4(a) and (b), the cavity mode frequency $f_{1,cav} > f_C$ and $f_{res} \approx f_{RO}$, resulting in a pronounced gain boost of the sidebands emerging around both the injected and un-injected mode at positive $f_m$ values. As the injection strength increases, the impact of the anti-guidance effect increases leading to a noticeable redshift of $f_{1,cav}$. This results in a decrease of $f_{res}$ and a shift of the gain peak towards lower frequency values, indicated with an arrow for both injected and un-injected signals in Fig. 4(a) and (b). For $\Delta f_{inj} = 1.2$ GHz, see Fig. 4(c) and (d), $f_{1,cav} < f_C$, and $f_{res} < 0$, resulting in a power increase at negative $f_m$ values for the sidebands around both the injected and un-injected modes. Besides the gain peaks due to the resonance frequency, the RO also has a strong influence by boosting the power of the sidebands for low $f_m$ values. Similarly to the case of negative detuning, increasing the injection strength enhances the anti-guidance effect, shifting the gain peak associated with the resonance frequency towards lower frequency values.

## IV. NUMERICAL MODEL AND SIMULATION

Looking at the dynamical behavior of semiconductor lasers, rate equation models have been shown, despite their simplicity, to qualitatively predict most if not all dynamical features reported experimentally [2,40]. Here, we use a phenomenological multimode model initially derived for solid-state lasers but adapted for semiconductor lasers [41]. We have extended this model to include single sideband injection to match our experimental scheme. The equations read as follows:

$$\frac{dF_1}{dt} = (1 + i\alpha)\left(g_1 N_1 - \frac{1-g_1}{2}\right) F_1 + \kappa_{inj} m(t) - i\Omega F_1, \quad (1)$$

$$\frac{dE_2}{dt} = (1 + i\alpha)\left(g_2 N_2 - \frac{1-g_2}{2}\right) E_2, \quad (2)$$

$$T\frac{dN_1}{dt} = P - N_1 - (1 + 2N_1)(g_1|E_1|^2 + g_2\beta|E_2|^2), \quad (3)$$

$$T\frac{dN_2}{dt} = P - N_2 - (1 + 2N_2)(g_2|E_2|^2 + g_1\beta|E_1|^2). \quad (4)$$

The first equation is written for $F_1(t) = E_1(t)e^{i\Omega t}$, i.e., the normalized electric field $E_1$ of mode 1 shifted in frequency by the normalized detuning $\Omega$ to make the equation autonomous. $E_2$ represents the normalized electrical field for mode 2, while $N_{1,2}$ are the normalized carrier population density. All variables and parameters are normalized by the photon lifetime $\tau_p$ as described in the original work [41]. To make the comparison with the experimental results easier, we use the non-normalized detuning $\Delta f_{inj} = \Omega/(2\pi\tau_p)$, where we chose $\tau_p = 3$ ps. The linewidth enhancement factor is set at $\alpha = 3$, and the normalized carrier lifetime to $T = 1250$. The pump parameter is defined as $P = (J - J_{th})/(2J_{th})$, with $J$ being the injected current and $J_{th}$ the threshold current. We set it at $P = 0.5$, corresponding to $J = 2J_{th}$. The mode coupling mechanism is driven by a parameter $\beta$ which models the cross-saturation. This parameter takes values between 0 and 1. A value of $\beta = 0$ describes two decoupled carrier pools whereas $\beta = 1$ refers to a single carrier pool shared between both modes. We set the cross-saturation parameter at $\beta = 0.95$. To tune the suppression ratio $\Delta P$ between the two modes of the MWL without injection, we fix the gain of the second mode at $g_2 = 0.995$, and adjust the modal gain $g_1$ of mode $E_1$ accordingly. Here, we set $g_1 = 0.954$ to achieve a suppression ratio of $\Delta P = 40$ dB, which is comparable to the one observed experimentally, see Fig. 1(b). The modulation term $\kappa_{\text{inj}} m(t)$ is introduced in Eq. (1), where $\kappa_{\text{inj}}$ is the injection rate, and $m(t) = 1 + A_m e^{j2\pi f_m t}$ is the single sideband signal. The modulation amplitude and frequency are represented by $A_m$ ($0 < A_m < 1$) and $f_m$, respectively. To qualitatively match our experimental conditions, we set the amplitude of the modulation to $A_m = 0.31$, resulting in about 10 dB power difference between the carrier and the sideband. When setting the detuning and injection strength at $\Delta f_{\text{inj}} = -6$ GHz and $\kappa_{\text{inj}} = 0.003$, respectively, and $f_m = 2.3$ GHz, the spectrum of the signal around the injected mode features the regenerated carrier ($C_{\text{reg}}$) and sideband ($SB_{\text{reg}}$) as well as the nonlinear sideband ($SB_{\text{NL}}$), see Fig. 5(a). As in the experiment, multiplied nonlinear sidebands $SB_{\text{mult,L}}$ and $SB_{\text{mult,R}}$ appear at a frequency offset from the un-injected mode equal to $f_m$, see Fig. 5(b). We observe that the two sidebands are symmetric and are consistent with a dual-sideband modulation, indicating that the initial single-sideband property is lost in the spectral multiplication, which is consistent with experimental observations. Figure. 5(c) and 5(d) display the power of the sidebands for both injected and un-injected modes with the same injection strength and detuning and while varying the modulation frequency from $f_m = -10$ GHz to $10$ GHz. As observed experimentally, the power of $SB_{\text{NL}}$ increases for $|f_m| \approx f_{\text{res}}$, while the power of $SB_{\text{reg}}$, increases for $f_m \approx f_{\text{res}}$, see Fig. 5(c). Similarly, the multiplied sidebands around the un-injected mode experience a power enhancement induced by the resonance frequency at the same frequency range as that of the injected signal. In addition, the power of the multiplied sidebands is significantly affected by a resonance at the $f_{\text{RO}} \approx 1.5$ GHz, see dashed lines in Fig. 5(d). We make the hypothesis that this feature is due to the substantial power difference, around 40 dB, between the injected and un-injected modes when no injection is applied. In other words, the un-injected mode is the driving force of the laser and is the only mode susceptible enough to such resonance. Hence, the RO frequency plays a more significant role in shaping the power of the multiplied sidebands emerging around the un-injected mode.

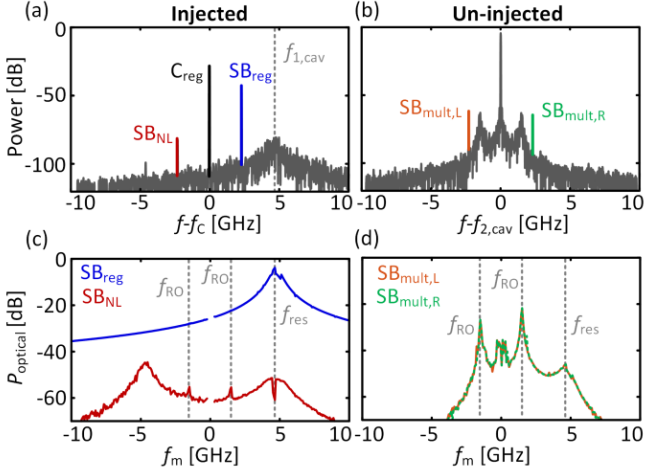

**Fig. 5.** Optical spectrum of the regenerated (a) and multiplied (b) signals around the injected and un-injected modes, respectively for $\Delta f_{\text{inj}} = -6$ GHz, $\kappa_{\text{inj}} = 0.003$, and $f_m = 2.3$ GHz. (c) Optical power of regenerated sidebands $SB_{\text{reg}}$ (dark blue trace) and $SB_{\text{NL}}$ (red trace) emerging around the injected mode and (d) of the multiplied sidebands $SB_{\text{mult,R}}$ (green trace) and $SB_{\text{mult,L}}$ (orange trace) emerging around the un-injected mode.

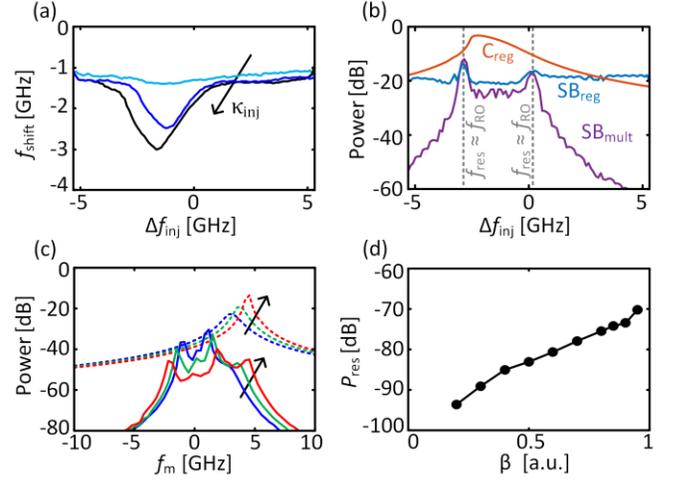

**Fig. 6.** (a) Frequency shift of $f_{\text{cav},1}$ induced by the anti-guidance effect at $\kappa_{\text{inj}} = 0.001$, $\kappa_{\text{inj}} = 0.003$, and $\kappa_{\text{inj}} = 0.005$ shown in light blue, dark blue and black respectively. (b) Power of $SB_{\text{reg}}$ (blue), $SB_{\text{mult}}$ (purple), and $C_{\text{reg}}$ (orange) at $f_m \approx f_{\text{res}}$, with $\kappa_{\text{inj}} = 0.005$, for detuning values ranging from $-5$ GHz to $5$ GHz. (c) Power of $SB_{\text{reg}}$ (dashed lines) and $SB_{\text{mult}}$ (solid lines) with $\kappa_{\text{inj}} = 0.005$ and $\Delta f_{\text{inj}} = -5$ GHz, for different pump current values: $J = 1.5 J_{\text{th}}$, $2 J_{\text{th}}$ and $3 J_{\text{th}}$, in blue, green, and red, respectively. (d) Power of the multiplied sidebands versus $\beta$ at $f_m \approx f_{\text{res}}$ and with $\kappa_{\text{inj}} = 0.005$ and $\Delta f_{\text{inj}} = 10.6$ GHz.

Next, we investigate the impact of injection parameters on the resonance frequency of the injected mode. Specifically, the cavity mode frequency, and consequently the resonance frequency, can be adjusted by varying the detuning and injection strength. In Fig. 6(a), we depict the resonance frequency shift $f_{\text{shift}} = f_{1,\text{cav}} - f_{1,\text{fr}}$ versus detuning for the injection strength of $\kappa_{\text{inj}} = 0.001$, $\kappa_{\text{inj}} = 0.003$, and $\kappa_{\text{inj}} = 0.005$. Due to the anti-guidance effect, the injected mode experiences a frequency shift toward lower frequency values, with the shift being more pronounced at negative detuning and higher injection strength values. In Fig. 6(b), we keep the injection strength constant at $\kappa_{\text{inj}} = 0.005$ and we analyze the impact of the detuning on the power of the regenerated and multiplied signals at $f_m \approx f_{\text{res}}$, because at this modulation frequency the power of $SB_{\text{reg}}$ and the combined power of the multiplied sidebands ($SB_{\text{mult}}$) is increased. We choose different detuning values, approximately from $-5$ GHz to $5$ GHz. The power of the regenerated carrier $C_{\text{reg}}$ exhibits a gain boost for the detuning values where $C_{\text{reg}}$ is close $f_{1,\text{cav}}$, see the orange trace in Fig. 6(b). The blue and purple traces in Fig. 6(b), represent the power of $SB_{reg}$ and $SB_{mult}$, respectively, which feature a gain boost for detuning values leading to $f_{res} \approx f_{RO}$, see the gray dashed lines in Fig. 6(b).

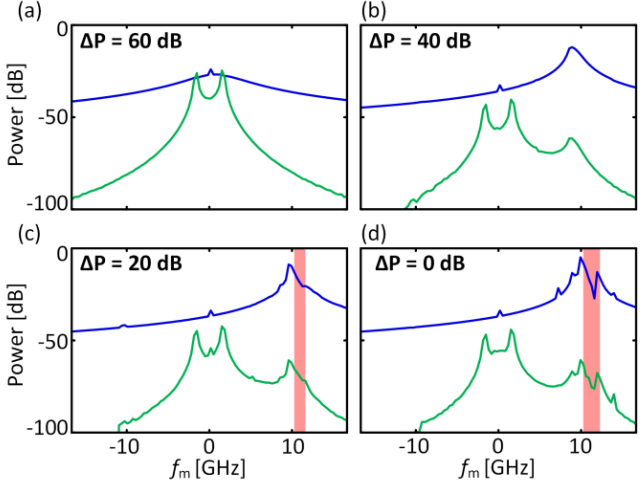

**Fig. 7.** Power of $SB_{reg}$ (blue) and $SB_{mult}$ (green) versus modulation frequency $f_m$ for fixed $\kappa_{inj} = 0.005$ and $\Delta f_{inj} = -10.6$ GHz. Pannels represents different values of $\Delta P$: (a) $\Delta P = 60$ dB, (b) $\Delta P = 40$ dB, (c) $\Delta P = 20$ dB, and (d) $\Delta P = 0$ dB. The red shaded areas show the regions where dynamical behavior is observed.

In the next step, we analyze the impact of the pump current on the power of the regenerated and multiplied signals. Figure. 6(c), represents the power of $SB_{reg}$ (dashed lines) and $SB_{mult}$ (solid lines) for a fixed $\kappa_{inj} = 0.005$ and $\Delta f_{inj} = -5$ GHz at three different pump current values: $P = 0.25$, $P = 0.5$, and $P = 1$, which, respectively, correspond to a bias current $J$ of 1.5, 2, and 3 times $J_{th}$. Since the power evolution of $SB_{NL}$ follows a similar trend as $SB_{reg}$ and features significantly less power, we refrain from showing it in the plot in Fig. 6(c). Increasing the bias current leads to the shift of the cavity mode toward higher frequencies due to the anti-guidance effect [37,38]. In addition, the RO frequency of the laser is proportional to the square root of the pump current. The increase in bias current leads to a rise in the RO frequency. As a result, higher bias currents shift the $f_m$ value at which the power increase of the sidebands occur. This comes with a reduction of the peak height, suggesting a higher damping of the RO.

Considering the complexity of the coupling mechanism between each pair of modes, represented here by the cross-saturation $\beta$, it is worthwhile to explore how this parameter affects the frequency response of the laser. To bring new insight into this question, we investigate the power evolution of the multiplied sidebands at $\kappa_{inj} = 0.005$ and $\Delta f_{inj} = 10.6$ GHz and for different values of the mode coupling parameter $\beta$. An important aspect to consider is that cross-saturation influences the power balance between the modes. To keep the same balance between the two modes of the laser, we need to carefully adjust the modal gain for each cross-saturation value [42]. Here we adjust the modal gain to maintain a 40 dB suppression ratio between the dominant and suppressed mode for different values of $\beta$ ranging from $\beta = 0$ to $\beta = 0.95$. When $\beta < 0.2$, the coupling between the injected and un-injected modes is remarkably poor and the two modes of the laser behave almost as independent lasers. In this scenario, the modulation of carrier density is not transferred to the un-injected mode, leading to the absence of sidebands around the un-injected modes. By increasing the cross-coupling, the two modes of the laser become more correlated and the sidebands start emerging around the un-injected mode. We compute the power of the multiplied sideband at $f_m \approx f_{res}$ for different mode-coupling values, see Fig. 6(d). At this frequency, the power of the multiplied sideband increases exponentially with the increase of $\beta$ parameter.

Finally, we analyze the impact of $\Delta P$ parameter on the power evolution of $SB_{reg}$ and $SB_{mult}$. This analysis is conducted with a fixed injection strength of $\kappa_{inj} = 0.005$ and a detuning of $\Delta f_{inj} = -10.6$ GHz, while varying $f_m$, see Fig. 7. To modify $\Delta P$, we set the modal gain of the injected mode to $g_1 = 0.85$, 0.954, 0.974, and 0.995 while keeping $g_2$ and $\beta$ fixed at $g_2 = 0.995$ and $\beta = 0.95$. These settings allow us to achieve $\Delta P$ values of 60 dB, 40 dB, 20 dB, and 0 dB respectively. When the power difference between the injected and un-injected modes is high as in Fig. 7(a) where $\Delta P = 60$ dB, $SB_{reg}$ and $SB_{mult}$ exhibit a slight amplification close to the RO frequency of the laser. In this scenario, the resonance frequency has no visible impact, as there is no noticeable gain boost observed around $f_m \approx f_{res}$, see Fig. 7(a). When the power difference between the injected and un-injected modes decreases, the gain boost induced by the resonance frequency becomes more pronounced and it is clearly visible in the power evolution of both $SB_{reg}$ and $SB_{mult}$ as shown in Fig. 7(b). When further reducing the power difference, we observe dynamic behavior emerging at the frequency range close to $f_{res}$. This is in agreement with the experimental observations, see Section 2 in Supplementary Materials. The dynamical regions are shown by red-shaded boxes in Fig. 7(c) and (d), where we observe, in particular, periodic oscillations that increase in complexity when reducing $\Delta P$, cf. Section 1 of the Supplementary Materials. It is worthwhile to highlight that the region exhibiting dynamic behavior expands proportionally with the increase in $\Delta P$. Our numerical investigations, therefore, suggest that the power difference between the injected and un-injected modes plays an important role in the power evolution of the regenerated and multiplied signals. However, the experimental confirmation needs a systematic investigation and remains to be explored in detail in future works.

## V. CONCLUSION

We have investigated the nonlinear response of an on-chip multi-wavelength laser under optical injection with single-sideband modulation. Through a combination of numerical simulations and experimental analyses, we have further analyzed the spectral multiplication phenomenon induced by the optical injection of SSB modulation into the MWL. This multiplication – a result of robust nonlinear coupling among different modes – is accompanied by asymmetric power of the sidebands with respect to the modulation frequency. Furthermore, our findings underscore the critical role played by the resonance frequency of the injected mode in determining the frequency response of the signal emerging around both the injected and un-injected modes. This dependency is effectively manipulated by adjusting the injection strength and detuning of the injected signal. We obtained a good agreement between our experimental observations and modeling based on a rather simple rate equation model including a cross-saturation between the two modes. We have highlighted that this cross-saturation parameter might have a leading role in shaping the response of the laser to optical injection. In addition, we highlight an important dependence of the dynamical behavior experienced by the laser's mode on the suppression ratio between the two modes of the laser. Overall, our study provides valuable insights into the intricate dynamics of on-chip multi-wavelength lasers under optical injection, paving the way toward precise control and optimization in future applications.

ACKNOWLEDGMENT

This work was supported by the European Research Council (ERC, Starting Grant COLOR'UP 948129, MV), the Research Foundation Flanders (FWO, grants 1530318N, G0G0319N, MV, and postdoctoral fellowship grant 1275924N, PMP), the METHUSALEM program of the Flemish Government (Vlaamse Overheid).

DATA AVAILABILITY STATEMENT

The experimental and numerical data produced in this study, as presented in this manuscript, have been deposited in the open-access digital repository Zenodo. They can be accessed at https://doi.org/10.5281/zenodo.13143636.

# Response of a multi-wavelength laser to single-sideband optical injection
## - Supplementary Material -

Shahab Abdollahi[*], Pablo Marin-Palomo, Martin Virte

*Brussels Photonics (B-PHOT), Vrije Universiteit Brussel, Pleinlaan 2, 1050 Brussel, Belgium*
[*] *Mohammadshahab.Abdollahi@vub.be*

### 1. Numerical analysis of MWL behavior under SSB injection when $\Delta P = 0$

To clarify the dynamical behaviors emerging in the red-shaded areas of Fig. 7(c) and (d) in the main paper, we present additional analysis in this section. The injection strength and detuning are similar to that of the main paper. The modulation frequency is set at $f_\mathrm{m} = 10.94$ GHz so that both modes show dynamical behavior and the laser operates within the red-shaded areas of Fig. 7(c) and (d). Figures S1(a–c) display the optical spectra and temporal evolution of the injected and un-injected modes at $f_\mathrm{m} = 10.94$ GHz and $\Delta P = 10$ dB, represented by red and blue traces, respectively. As shown, periodic oscillation is the dominant behavior. We examined other values of $f_\mathrm{m}$ within this dynamical region and consistently observed the same behavior. Similarly, to analyze the dynamics at $\Delta P = 0$ dB, Figures S1(d–f) illustrate the optical spectra and temporal evolution of the modes at $f_\mathrm{m} = 10.94$ GHz. In this case, higher-order periodic oscillations emerge around both the injected and un-injected modes. This highlights the increasing complexity of the system as the power levels between the modes become similar. It is worth mentioning that since the injection strength used for the simulation is relatively low, the dynamical region is limited to a few gigahertz. By increasing the injection strength, the region where dynamical behavior is observed would expand. However, this also increases the likelihood of suppressing the un-injected mode, which is dominant prior to injection.

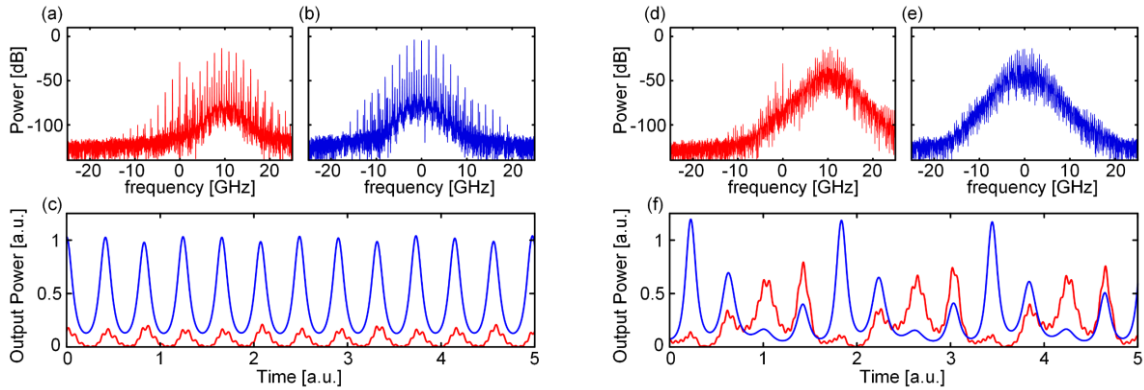

**Figure S1.** Spectral and temporal evolution of the injected and un-injected modes. Spectral evolution of the injected (a) and un-injected modes (b) where $g_2 = 0.974$ ($\Delta P = 20$ dB) and $f_\mathrm{m} = 10.94$ GHz. (c) corresponding temporal evolution is depicted by red and blue lines for both injected and un-injected modes respectively. Spectral evolution of the injected (d) and un-injected modes (e) where $g_2 = 0.995$ ($\Delta P = 0$ dB) and $f_\mathrm{m} = 10.94$ GHz. (f) corresponding temporal evolution is depicted by red and blue lines for both injected and un-injected modes respectively.

## 2. Experimental analysis of MWL behavior under SSB injection when ΔP = 0

In the main paper, we experimentally show the response of the injected and un-injected modes of the MWL to SSB signal injection when the injected mode is strongly suppressed prior to the injection. In this section of the SI, we explore the laser's response to SSB injection when both modes have identical power levels. To balance the power between the injected and un-injected modes of the laser, we applied a current of 1.8 mA to DBR1, while DBR2 remained unbiased. All other laser parameters were kept consistent with those in the main paper. Applying current to DBR1 triggers emission from both cavities with nearly equal power, see Fig. S2. After balancing the power between the modes, the SSB signal is injected around the mode at 1536.88 nm, as indicated by an arrow in Fig S2. Due to strong mode competition among the various modes of the MWL, the mode emitting at 1547.78 nm became completely suppressed following the injection. Consequently, we focused our analysis on the un-injected mode at 1548.61 nm, also marked by an arrow. In this scenario, the injection strength plays a critical role because excessive injection strength can easily suppress all un-injected modes. To prevent this suppression, we used a relatively low injection strength of $\kappa_{inj} = -2$ dB with the detuning set to $\Delta f_{inj} = 2$ GHz. At this detuning value, the cavity mode appears on the left side of the carrier, as shown by a gray dashed line in Fig. S3(a). The sideband with $P_m = -10$ dB is swept from negative to positive modulation frequency values. Figures. S3(a) and (b) show the optical spectra of the injected and un-injected modes at $f_m = -3.12$ GHz. The second row of Fig. S3 illustrates the power evolution of the regenerated and nonlinear sidebands around the injected mode in Fig. S3(c), and the left and right multiplied sidebands around the un-injected mode in Fig. S3(d). keeping the balanced power between the injected and un-injected modes results in stronger signal multiplication around the un-injected mode. This occurs because the injected signal experiences greater amplification as it passes through the cavity resonance frequency, where the injected mode exhibits a higher gain. As a result, a more powerful signal is multiplied around the un-injected mode. This enhanced signal multiplication is accompanied by increased dynamical behavior emerging around both the injected and un-injected modes, as indicated by the red-shaded areas in Fig. S3(c) and (d). This phenomenon is in agreement with our simulation result shown in Fig. 7(d) of the main paper.

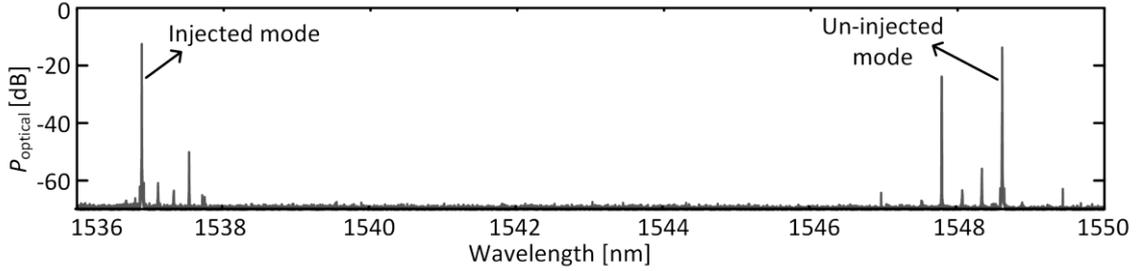

**Figure S2.** Optical spectrum of the MWL prior to injection, when the injected and un-injected modes emit with balanced power. The injected and un-injected modes, indicated by arrows, are later analyzed once the injection is applied.

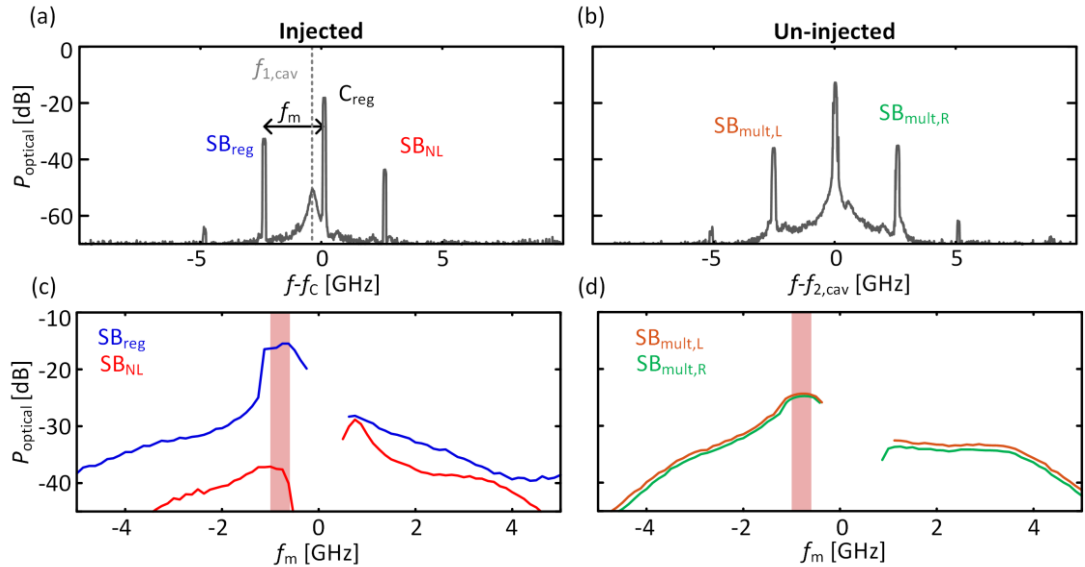

**Figure S3.** Optical spectra of the regenerated (a) and multiplied (b) signals around the injected and un-injected modes, respectively for $\Delta f_{inj} = 2$ GHz, $\kappa_{inj} = -2$ dB, and $f_m = -3.12$ GHz. (c) Power evolution of $SB_{reg}$ (dark blue trace) and $SB_{NL}$ (red trace) emerging around the injected mode and (d) of $SB_{mult,R}$ (green trace) and $SB_{mult,L}$ (orange trace) emerging around the un-injected mode. The red shaded areas indicate the regions with dynamical behavior.

In the next experiment, we increased the injection strength to $\kappa_{inj} = 2$ dB, with the detuning set at $\Delta f_{inj} = -0.7$ GHz. Figure. S4(a) and (b) display the optical spectra of the injected and un-injected modes when $f_m = -3.12$ GHz. The strong injection causes the un-injected mode to be significantly suppressed by more than 30 dB, as shown in Fig. S4(b). This is due to the mode competition inside the shared gain medium and the fact that optical injection moves the balance toward the injected mode. Consequently, the un-injected mode exhibits no signal multiplication due to its strong suppression. Figure. S4(c) illustrates the power evolution of the regenerated and nonlinear sidebands, while the un-injected mode leaves no trace in the power evolution. In the case of high injection strengths, we also observe a region featuring high-order dynamics as shown by the red-shaded area in Fig. S4(c).

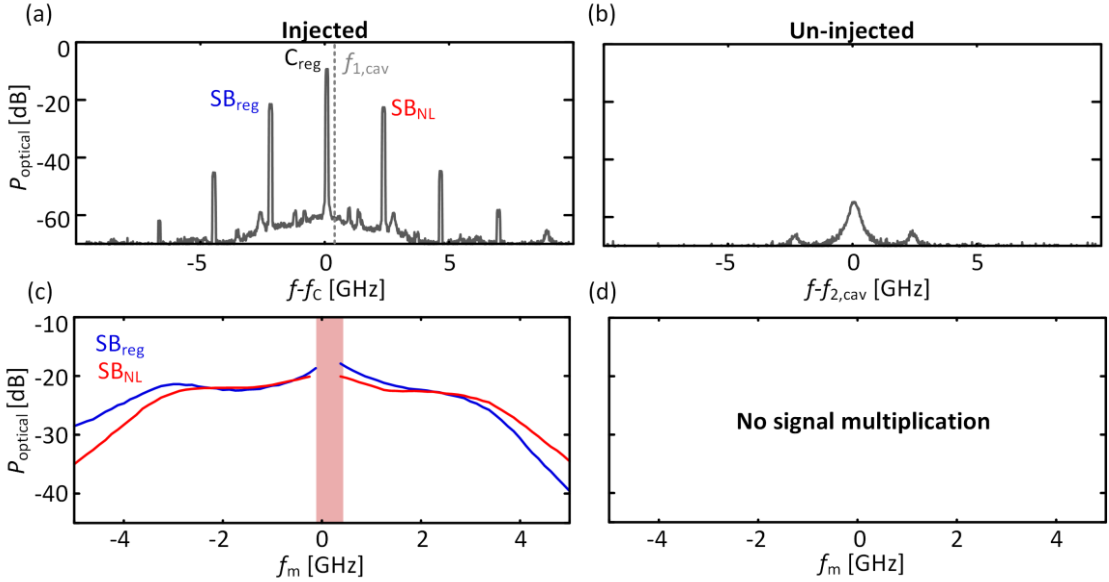

**Figure S4.** Optical spectra of the injected (a) and un-injected (b) modes when SSB signal with $\Delta f_{inj} = -0.7$ GHz, $\kappa_{inj} = 2$ dB, and $f_m = -3.12$ GHz is injected. (c) Power evolution of SB$_{reg}$ (dark blue trace) and SB$_{NL}$ (red trace) emerging around the injected mode, while (d) no signal is multiplied around the un-injected mode. The red-shaded area indicates the regions with dynamical behavior.